\documentclass{IEEEtran}
\usepackage{epsfig,makeidx,color}
\usepackage{graphicx}
\usepackage{amsbsy}
\usepackage{amsmath}
\usepackage{amssymb}
\usepackage{euscript}
\usepackage{citesort}

\newtheorem{myproperty}{\bf Property}

\newtheorem{myremark}{\bf Remark}

\newtheorem{mydefinition}{\bf Definition}
\newtheorem{myproposition}{\bf Proposition}

\def\beq{\begin{equation}}
\def\eeq{\end{equation}}
\def\vec{\mbox{vec}}

\begin{document}
\title{On Fast-Decodable Space--Time Block Codes}
\author{Ezio Biglieri $\qquad$ Yi Hong  $\qquad$ Emanuele Viterbo
\thanks{\scriptsize  Ezio Biglieri is with Departament de Tecnologies
de la Informaci\'o i les Comunicacions, Universitat Pompeu Fabra
(DTIC-UPF), Barcelona, Spain. E-mail: $\tt e.biglieri$@$\tt
ieee.org$. Yi Hong was with DEIS - Universit\`{a} della Calabria,
and is now with Institute of Advanced Telecom., University of
Wales, Swansea, Singleton Park, SA2 8PP, UK. Email: $\tt
y.hong$@$\tt swansea.ac.uk$. Emanuele Viterbo is with DEIS -
Universit\`{a} della Calabria, via P. Bucci, 42/C, 87036 Rende
(CS), Italy.  E-mail: $\tt viterbo$@$\tt deis.unical.it$. This
work was supported by the STREP project No. IST-026905 (MASCOT)
within the Sixth Framework Programme of the European Commission.
Ezio Biglieri's work was also supported by Sequans Communications,
Paris, France.}}
\maketitle
\begin{abstract}
We focus on full-rate, fast-decodable space--time block codes
(STBCs) for $2\times 2$ and $4\times 2$ multiple-input
multiple-output (MIMO) transmission. We first derive conditions
and design criteria for reduced-complexity maximum-likelihood
decodable $2\times 2$ STBCs, and we apply them to two families of
codes that were recently discovered.  Next, we derive a novel
reduced-complexity $4\times 2$ STBC, and show that it outperforms
all previously known codes with certain constellations.
\end{abstract}


\section{Introduction}

In 1998, Alamouti~\cite{Alamouti} invented a remarkable scheme for
multiple-input multiple-output (MIMO) transmission using two
transmit antennas and admitting a low-complexity
maximum-likelihood (ML) decoder. Space--time block codes (STBCs)
using more than two transmit antennas were designed
in~\cite{TarokhSTBC}. For such codes, ML decoding is achieved in a
simple way, but, while they can achieve maximum diversity
gain~\cite{Guey,Tarokh98}, their transmission rate is reduced.
The {\em quasi-orthogonal} STBCs in~\cite{JafarSTBC} can support a
transmission rate larger than orthogonal STBCs, but at the price
of a smaller diversity gain. Using algebraic number theory and
cyclic division algebras, algebraic STBCs can be designed to
achieve full rate and full diversity, but at the price of a higher
decoding complexity.

Recently, a family of $2\times 2$ {\em twisted space--time
transmit diversity} STBCs, having full rate and full diversity,
was proposed in~\cite{TirIT,Tirkkonen,Hottinen,Hottinenbook}.
These codes were recently rediscovered in~\cite{paredes}, whose
authors also pointed out that they enable reduced-complexity ML
decoding (see {\em infra} for a definition of decoding
complexity). Independently, the same codes were found
in~\cite{FitzIsit07}. More recently, another family of full-rate,
full-diversity, fast-decodable $2\times 2$ codes for MIMO was
proposed in~\cite{Sari}.

Empirical evidence seems to show that the constraint of simplified
ML decoding does not entail substantial performance loss. To
substantiate the above claim, the present paper provides a unified
view of the fast-decodable STBCs
in~\cite{TirIT,Tirkkonen,Hottinen,paredes,FitzIsit07,Sari} for
$2\times 2$ MIMO. We show that all these codes allow the same
low-complexity ML decoding procedure, which we specialize in the
form of a sphere-decoder (SD) search~\cite{Caire,SE,VB1,VB2}. We
also derive general design criteria for full-rate, fast-decodable
STBCs, and we use it to design a family of $4\times 2$ codes based
on a combination of algebraic and quasi-orthogonal structures. In
this case, the full-diversity assumption is dropped in favor of
simplified maximum-likelihood decoding. Within this family, we
exhibit a code that outperforms all previously proposed $4\times
2$ STBCs for $4$-QAM signal constellation.

The balance of this paper is organized as follows.
Section~\ref{sec:sysmod} introduces system model and code design
criteria. In Section~\ref{sec:GeneralDecoding}, we present the
concept of the fast-decodability of STBCs. In
Section~\ref{sec:STBC22} we review two families of fast-decodable
$2\times 2$ STBCs recently appeared in the literature, and we show
how both of them enable a reduced-complexity ML decoding procedure.
In Section~\ref{sec:newSTBC42}, we propose fast-decodable $4\times
2$ STBCs, and we show the corresponding ML decoding complexity.
Finally, conclusions are drawn in Section~\ref{Concl}.

{\em Notations}: Boldface letters are used for column vectors, and
capital boldface letters for matrices. Superscripts $^T$,
$^{\dagger}$, and $^*$ denote transposition, Hermitian
transposition, and complex conjugation, respectively. $\mathbb{Z}$,
$\mathbb{C}$, and $\mathbb{Z}[j]$ denote the ring of rational
integers, the field of complex numbers, and the ring of Gaussian
integers, respectively, where $j^2 =-1$. Also, $\mathbf{I}_{n}$
denotes the $n\times n$ identity matrix, and $\mathbf{0}_{m\times
n}$ denotes the $m\times n$ matrix all of whose elements are $0$.

Given a complex number $x$, we define the $\tilde{(\cdot)}$
operator from $\mathbb{C}$ to $\mathbb{R}^2$ as $ \tilde{x}
\triangleq [\Re(x), \Im(x)]^T, $ where $\Re{(\cdot)}$ and
$\Im{(\cdot)}$ denote real and imaginary parts. The
$\tilde{(\cdot)}$ operator can be extended to complex vectors
$\mathbf{x}= [x_1, \ldots x_n] \in \mathbb{C}^n$:
\[
\tilde{\mathbf{x}} \triangleq 
 [\Re(x_1), \Im(x_1), \ldots, \Re(x_n), \Im(x_n)]^T
\]
Given a complex number $x$, the $\check{(\cdot)}$ operator from
$\mathbb{C}$ to $\mathbb{R}^{2\times 2}$ is defined by
\[ \check{x} \triangleq \left[ \begin{array}{cc} \Re(x) & -\Im(x) \\  \Im(x) & \Re(x)
\end{array}\right]\]
The $\check{(\cdot)}$ operator can be similarly extended to
$n\times n$ matrices by applying it to all the entries, which
yields $2n\times 2n$ real matrices. The following relations hold:
$
\widetilde{\mathbf{Ax}} = \check{\mathbf{A}} \tilde{\mathbf{x}}
$
and
$
 \mathbf{A}=\mathbf{B}\mathbf{C} \Longrightarrow
 \check{\mathbf{A}}=\check{\mathbf{B}}\check{\mathbf{C}}
$. Given a complex number $x$, we define the $\bar{\check{(\cdot)}}$
operator from $\mathbb{C}$ to ${\mathbb{R}}^{2\times 2}$ as
\[
 \bar{\check{x}} \triangleq \left[ \begin{array}{cc} -\Re(x) & -\Im(x) \\  -\Im(x) & \Re(x)
\end{array}\right]
\]
The following relation holds:
\[
 \widetilde{xy^{*}}\triangleq \bar{\check{x}}\cdot \tilde{y}
\]
The ${\rm vec}(\cdot)$ operator stacks the $m$ column vectors of a
$n\times m$ complex matrix into a $mn$ complex column vector. The
$\Vert \cdot \Vert$ operation denotes the Euclidean norm of a
vector. Finally, the Hermitian inner product of two complex column
vectors $\mathbf{a}$ and $\mathbf{b}$ is denoted by $\langle
\mathbf{a}, \mathbf{b}\rangle \triangleq \mathbf{a}^{T}
\mathbf{b}^{*}$. Note also that if
$\langle\mathbf{a},\mathbf{b}\rangle=0$, then
$\langle\tilde{\mathbf{a}},\tilde{\mathbf{b}}\rangle=0$.

\section{System Model and Code Design Criteria}\label{sec:sysmod}
We consider a $n_r\times n_t$ MIMO transmission over a
block-fading channel. The received signal matrix ${\mathbf{Y}} \in
\mathbb{C}^{n_r\times T}$ is
\begin{equation}\label{systemmodel}
{\mathbf{Y}}={\mathbf{H}}{\mathbf{X}}+{\mathbf{N}}
\end{equation}
where $\mathbf{X} \in \mathbb{C}^{n_t\times T}$ is the codeword
matrix, transmitted over $T$ channel uses. Moreover,
${\mathbf{N}}\in \mathbb{C}^{n_r\times T}$ is a complex white
Gaussian noise with i.i.d.\ entries $\sim {\EuScript
N}_{\mathbb{C}} (0,N_0)$, and $\mathbf{H}=[h_{i\ell}]\in
\mathbb{C}^{n_r\times n_t}$ is the channel matrix, assumed to
remain constant during the transmission of a codeword,  and to
take on independent values from codeword to codeword. The elements
of $\mathbf{H}$ are assumed to be i.i.d.\ circularly symmetric
Gaussian random variables $\sim {\EuScript N}_{\mathbb{C}} (0,1)$.
The realization of $\bf H$ is assumed to be known at the receiver,
but not at the transmitter. The following definitions are relevant
here:
\begin{mydefinition}
{\bf\em (Code rate)} Let $\kappa$ be the number of independent
information symbols per codeword, drawn from a complex constellation
${\EuScript S}$. The code rate of a STBC is defined as $R =
\kappa/T$ symbols per channel use. If $\kappa=n_r T$, the STBC is
said to have {\em full rate}. $\hfill\square$
\end{mydefinition}

Consider ML decoding. This consists of finding the code matrix
that achieves the minimum of the squared Frobenius norm $m({\bf
X})\triangleq\| {\bf Y-HX} \|^2$.
\begin{mydefinition}
 {\bf\em (Decoding Complexity)} The ML decoding complexity
is defined as the minimum number of values of $m(\bf X)$ that
should be computed in ML decoding. This number cannot exceed
$M^\kappa$, with $M =|{\EuScript S}|$, the complexity of the
exhaustive-search ML decoder. $\hfill\square$
\end{mydefinition}

Consider two codewords $\mathbf{X}$ and $\widehat{\mathbf{X}} \ne
{\bf X}$. Let $r$ denote the minimum rank of the matrix
$\mathbf{X} -\widehat{\mathbf{X}}$,  and $\delta$ the {\em product
distance}, i.e., the product of non-zero eigenvalues of the
codeword distance matrix $\mathbf{E}\triangleq (\mathbf{X}
-\widehat{\mathbf{X}})(\mathbf{X} -\widehat{\mathbf{X}})^\dagger$.
The error probability of a STBC is upper bounded by the following
union bound,
\begin{equation}
P(e) \leq \frac{1}{M^\kappa}\sum_{r}\sum_{\delta} A(r,\delta)
P(r,\delta) \label{UB}
\end{equation}
where $P(r,\delta)$ denotes the pairwise error probability (PEP) of
the codeword differences with rank $r$ and product distance
$\delta$, and $A(r,\delta)$ the associated multiplicity. In
\cite{Tarokh98}, the ``rank-and-determinant criterion'' (RDC) was
proposed to maximize both the minimum rank $r$ and the {\em minimum
determinant} $\delta_{\min} \triangleq \underset{\mathbf{X}
\ne\widehat{\mathbf{X}}} {\min}\det\left( \mathbf{E} \right)$. For a
full-diversity STBC, i.e., $r=n_t$ for all $\mathbf{E}$ matrices,
this criterion yields {\em diversity gain} $n_r n_t$ and {\em coding
gain} $\left( \delta_{\min}
\right)^{\frac{1}{n_t}}$~\cite{Tarokh98}. For STBC with
$\delta_{\min}=0$, and hence without full diversity, one should
minimize $A(r,\delta)$ with $r\leq n_t$.

\subsection{Linear codes, and Codes with the Alamouti structure}
{\em Linear} STBCs are especially relevant in our context, because
they admit ML sphere decoding.
\begin{mydefinition}
 {\bf\em (Linear STBC)} A STBC carrying $\kappa$ symbols $\mathbf{s}=[s_1,\ldots,s_\kappa]$ is said
 to be {\em (real) linear} if we can write
$
 \widetilde{\vec(\mathbf{X})} = \mathbb{G}\tilde{\mathbf{s}}
$
for some $\mathbb{G} \in \mathbb{R}^{2n_t T\times 2 \kappa}$. The
matrix $\mathbb{G}$ is called the {\em (real) generator matrix} of
the linear code. If a complex matrix $\mathbf{G}\in
\mathbb{C}^{n_t T\times \kappa}$ exists such that $\mathbb{G} =
\check{\mathbf{G}}$, then we can write
$
 \vec(\mathbf{X}) = \mathbf{G}\mathbf{s}
$
which identifies a {\em complex linear} STBC, with $\mathbf{G}$
its {\em complex generator matrix}. $\hfill\square$
\end{mydefinition}
\begin{mydefinition}
{\bf\em (Cubic shaping)} For a linear STBC, if its real generator
matrix $\mathbb{G}$ is an orthogonal matrix satisfying
$\mathbb{G}^{\dagger}\mathbb{G} = \mathbf{I}_{2\kappa}$, then we
say that the STBC {\em has cubic shaping} (see~\cite{FNT} for the
significance of cubic shaping). $\hfill\square$
\end{mydefinition}

Linear STBCS admit the canonical decomposition
 \beq\label{candec}
 {\bf X} = \sum_{\ell=1}^\kappa (a_\ell {\bf A}_\ell +  j b_\ell {\bf B}_\ell )
 \eeq
 where $a_\ell$ and $b_\ell$ are the real and imaginary part
 of $s_\ell$, respectively, and ${\bf A}_\ell, {\bf B}_\ell$,
$\ell=1, \ldots, \kappa$, are
 $n_t\times T$ (generally complex) matrices. With this
 decomposition,~(\ref{systemmodel})
 can be rewritten using only real quantities:
\begin{equation}\label{basic}
\widetilde{\vec\, ({\bf Y})} = {\mathbb F}{\tilde{\bf s}} +
\widetilde{\vec \,({\bf N})}
\end{equation}
where
\begin{eqnarray*}
  {\mathbb{F}} &\triangleq& [ \widetilde{\vec({\bf HA}_1)},  \widetilde{\vec({\bf
  HB}_1)}, \cdots, \widetilde{\vec({\bf HB}_\kappa)}]\\ &=&
  {\rm diag}(\check{\mathbf{H}},\ldots,\check{\mathbf{H}})\, \mathbb{G}
\end{eqnarray*}
and
$\mathbb{G}=[\widetilde{\vec(\mathbf{A}_1)},\ldots,\widetilde{\vec(\mathbf{B}_\kappa)}]$.
Note that the $\mathbb F$ matrix depends on $\mathbf H$. With
complex linear STBC, we may use only complex quantities:
 \beq\label{complex}
  \vec\, ({\bf Y}) = {\mathbf F}{\bf s} + \vec \,({\bf N})
  \eeq
  where now
\begin{eqnarray}
  \mathbf{F} &\triangleq& [\vec(\mathbf{HA}_1),\vec(\mathbf{HB}_1),\ldots,\vec(\mathbf{HB}_\kappa)]
 \nonumber \\ &=& {\rm diag}(\mathbf{H},\ldots,\mathbf{H})\, \mathbf{G}
 \label{complexF}
   \end{eqnarray}
with
$\mathbf{G}=[\vec(\mathbf{A}_1),\ldots,\vec(\mathbf{B}_\kappa)]$,
$\check{\mathbf{G}}=\mathbb{G}$, and
$\check{\mathbf{F}}=\mathbb{F}$.

\begin{mydefinition}
{\bf\em (Alamouti structure)}~~ We say that a STBC {\em has the
Alamouti structure} if
\begin{equation}\label{Alamouti_Structure}
 \mathbf{X} =
    \left[
\begin{array}{cr}
 \alpha s_1 & -\beta s^{*}_2 \\
 \alpha s_2 & \beta s^{*}_1
\end{array}%
\right]
\end{equation}
where $s_i\in {\mathbb{C}}$ with $i=1,2$, and $\alpha, \beta \in
\mathbb{C}$, $\vert \alpha \vert^2 = \vert\beta \vert^2$, and
$\vert \alpha \vert^2 + \vert\beta \vert^2 = 1$. $\hfill\square$
\end{mydefinition}

From the definition of linear codes, we have
 \begin{eqnarray}\label{generator1}
 {\mathbb G} = \left[\begin{array}{cc}
 \check{\alpha} & \check{0}\\
 \check{0} & \check{\alpha}\\
 \check{0} & \bar{\check{\beta}}\\
 -\bar{\check{\beta}} & \check{0}
\end{array} \right]
\end{eqnarray}
and can see, by direct calculation, that ${\mathbb G}^T{\mathbb
G}={\bf I}_4$, which implies the cubic shaping of these STBCs.
Moreover, given ${\bf H} = [h_{ij}] \in {\mathbb C}^{2\times 2}$
and ${\bf Y} = [y_{ij}] \in {\mathbb C}^{2\times 2}$, let us
define
\begin{eqnarray}
 \mathbf{y} \triangleq [y_{11},y_{21},y^{*}_{12},y^{*}_{22}]^T
 ~~~~\mathbf{n} \triangleq [n_{11},n_{21},n^{*}_{12},n^{*}_{22}]^T
\end{eqnarray}
where the last two elements of the vectorized matrices are
conjugated. We can write (\ref{systemmodel}) as
\begin{eqnarray}\label{conjugatedsystemmodel}
\mathbf{y} = \mathbf{F}^{(*)} \mathbf{s} + \mathbf{n}
\end{eqnarray}
where
\begin{eqnarray}\label{F}
\mathbf{F}^{(*)}\triangleq [\mathbf{f}_1\vert\mathbf{f}_2]=\left[
\begin{array}{cc}
  {\alpha h_{11}} &  {\alpha h_{12}}\\
  {\alpha h_{21}} &  {\alpha h_{22}}\\
  {\beta^{*} h^{*}_{12}} &  {-\beta^{*} h^{*}_{11}}\\
  {\beta^{*} h^{*}_{22}} &  {-\beta^{*} h^{*}_{21}}
\end{array} \right]
\end{eqnarray}
and
\[
{\mathbb F}\triangleq
\check{\mathbf{F}}^{(*)}=[\check{\mathbf{f}}_1
\vert\check{\mathbf{f}}_2]
\]
%
Note that $\mathbf{F}^{(*)}$ has its last two rows conjugated. In
complex notations, multiplication of $\bf y$ at the receiver by
$\left(\mathbf{F}^{(*)}\right)^{\dagger}$ is equivalent to matched
filtering. Direct calculation shows that, for codes with the
Alamouti structure,
\begin{equation}\label{Alamoutistructure}
{\mathbf{F}^{(*)}}^{\dagger} \mathbf{F}^{(*)}= {\bf I}_2~~~~{\text
i.e.,}~~~~\langle \mathbf{f}_1, \mathbf{f}_2 \rangle = 0
\end{equation}
and hence ML decoding can be done symbol-by-symbol, which, under
our definition, yields complexity $2M$.

\section{Fast decoding with QR decomposition}\label{sec:GeneralDecoding}
Consider a linear STBC carrying $\kappa$ independent QAM information
symbols. Following (\ref{complex}), at the receiver, the SD
algorithm can be used to conduct ML decoding based on QR
decomposition of matrix $\bf F$~\cite{Caire}:
$\mathbf{F}=\mathbf{QR}$, where $\mathbf{Q}\in
\mathbb{C}^{\kappa\times \kappa}$ is unitary, and $\mathbf{R}\in
\mathbb{C}^{\kappa\times \kappa}$ is upper-triangular. The ML
decoder minimizes $\| \mathbf{Q}^{\dagger} {\vec({\bf Y})} -
{\mathbf R}{\bf s} \|$. If we write
\[
\mathbf{F} = [\mathbf{f}_1 \mid \mathbf{f}_2 \mid \ldots \mid
\mathbf{f}_{\kappa}]  \in\mathbb{C}^{\kappa\times \kappa}
\]
then the matrices $\bf Q$ and $\bf R$ have the general form
\[
\mathbf{Q} = \left[\mathbf{e}_1\mid\mathbf{e}_2\mid
\ldots\mid\mathbf{e}_{\kappa}\right]
 \]
 and
\[
\mathbf{R} =
\left[
\begin{array}{cccc}
 \Vert \mathbf{d}_1\Vert & \langle  \mathbf{f}_2, \mathbf{e}_1\rangle &
 \cdots & \langle  \mathbf{f}_{\kappa}, \mathbf{e}_1\rangle\\
  0 & \Vert \mathbf{d}_2\Vert & \ddots &
  \langle  \mathbf{f}_{\kappa}, \mathbf{e}_2\rangle\\
  0 & 0 & \ddots & \vdots\\
  0 & 0 & 0  &\Vert \mathbf{d}_{\kappa} \Vert
\end{array}%
\right]
\]
where
\begin{eqnarray*}
\mathbf{d}_1 &=&  \mathbf{f}_1 \qquad\qquad\qquad\qquad \mathbf{e}_1
= \frac{\mathbf{d}_1}{\Vert \mathbf{d}_1 \Vert}=
\frac{\mathbf{f}_1}{\Vert \mathbf{f}_1 \Vert}\nonumber\\\mathbf{d}_i
&=&  \mathbf{f}_i - \sum^{i-1}_{j=1} {\rm Proj}_{\mathbf{e}_j}
\mathbf{f}_i \qquad\, \mathbf{e}_i = \frac{\mathbf{d}_i}{\Vert
\mathbf{d}_i \Vert},~~ i=2,\cdots,\kappa
\end{eqnarray*}
and $\rm{Proj}_{\mathbf{u}} \mathbf{v} \triangleq \frac{\langle
\mathbf{v},\mathbf{u} \rangle} {\langle \mathbf{u},\mathbf{u}
\rangle}\mathbf{u}$. This formulation of the QR decomposition
coincides with the Gram-Schmidt procedure applied to the column
vectors of $\mathbf{F}$. It was pointed out in~\cite{Caire} that
the search procedure of a SD can be visualized as a bounded tree
search. If a {\it standard} SD is used for the above STBC, we have
$\kappa$ levels of the complex SD tree, where the worst-case
computation complexity is $M^\kappa$. However, zeros appearing
among the entries of $\mathbf R$ can lead to simplified SD, as
discussed in the following.

If the condition
\begin{equation} \label{fast-dec-condition}
\langle \mathbf{f}_2, \mathbf{e}_i\rangle = \langle \mathbf{f}_3,
\mathbf{e}_i\rangle = \cdots = \langle \mathbf{f}_{k'},
\mathbf{e}_i\rangle = 0
\end{equation}
is satisfied for $i=1,\ldots, k'-1$  and for some  $k'\leq \kappa$,
then $k'$ levels can be removed from the complex SD tree, and we can
employ a $(\kappa-k')$-dimensional complex SD. In it, we first
estimate the partial vector $(s_{k'+1}, \ldots, s_{\kappa})$. For
every such vector (there are $M^{\kappa-k'}$ of them), a linear ML
decoding, of complexity $k'M$, is used to choose $s_1,\ldots,s_{k'}$
so as to minimize the total ML metric. Hence, the worst-case
decoding complexity is $k' M^{(\kappa-k'+1)}$. The components $s_i$
should be sorted in order to maximize $k'$.

Analysis of the structure of the matrix $\mathbf R$ yields the
following observation:

Zero entries of $\mathbf R$, besides those
in~(\ref{fast-dec-condition}), lead to faster metric computations in
the relevant SD branches, but not to a reduction of the number of
branches. We conclude this Section with the following:
\begin{mydefinition}\label{Def:fastdecSTBC}
{\bf\em (Fast-decodable STBCs)}~~A linear STBC allows fast ML
decoding if (\ref{fast-dec-condition}) is satisfied, yielding a
complexity of the order of $k' M^{\kappa-k'+1}$. $\hfill\square$
\end{mydefinition}

\section{Fast-decodable codes for $2\times 2$ MIMO, and ML decoding}\label{sec:STBC22}

Consider now full-rate $(R=2)$ and full-diversity fast-decodable
$2\times 2$ STBCs, i.e., with $\kappa=4$~symbols/codeword and
$r=n_t$. Here we examine two families of $2\times 2$ full-rate,
full-diversity fast-decodable STBCs, endowed with the following
structure:
\begin{equation}
\label{transmitmatrix} {\bf X} = {\bf X}_{1,2}(s_1,s_2) + {\bf
X}_{3,4}(s_3,s_4)
\end{equation}
where the first (resp., second) component code encodes symbols
$s_1,s_2$ (resp., $s_3,s_4$).

{\bf Family I}:~~In this family of fast-decodable STBCs,
independently derived in~\cite{Tirkkonen,paredes,FitzIsit07},
$\mathbf{X}_{1,2}(s_1,s_2)$ has the Alamouti structure
\cite{Alamouti} with $\alpha = \beta = 1$ and
$\mathbf{X}_{3,4}(s_3,s_4)$ is chosen as follows: let
\begin{equation}\label{TUZ}
\mathbf{T} \triangleq
    \left[
\begin{array}{cc}
1 & 0 \\
0 & -1
\end{array}%
\right]~~{\rm and}~~
 \left[
\begin{array}{c}
z_1  \\
z_2
\end{array}%
\right] = \mathbf{U}\left[
\begin{array}{c}
s_3  \\
s_4
\end{array}%
\right]
\end{equation}
where $z_1, z_2 \in \mathbb{C}$, and $\mathbf{U}\in
\mathbb{C}^{2\times 2}$ is the unitary matrix
\[
 \mathbf{U} = \left[
\begin{array}{cc}
\varphi_1 & -\varphi^{*}_2 \\
\varphi_2 & \varphi^{*}_1
\end{array}%
\right]
\]
with $\vert \varphi_1\vert^{2}+ \vert \varphi_2 \vert^{2} = 1$. We
have
\begin{eqnarray}\label{XbFirstFamily}
    \lefteqn{\mathbf{X}_{3,4}(s_3,s_4) = \mathbf{T}
    \left[
\begin{array}{cc}
z_1 & -z^{*}_2 \\
z_2 & z^{*}_1
\end{array}%
\right]} \\  &=& \mathbf{T} \left[
\begin{array}{cc}
\varphi_1 s_3 -\varphi^{*}_2 s_4 & -(\varphi_2 s_3 + \varphi^{*}_1 s_4)^{*}\\
\varphi_2 s_3 + \varphi^{*}_1 s_4 & (\varphi_1 s_3 -\varphi^{*}_2 s_4)^{*}
\end{array}%
\right] \nonumber
\end{eqnarray}
which has the Alamouti structure~(\ref{Alamouti_Structure}).
Vectorizing, and separating real and imaginary parts of the matrix
$\mathbf{X}$, we obtain
\[
\widetilde{{\rm vec}(\mathbf{X})} = \mathbb{G}\, [
\tilde{s}_1,\tilde{s}_2,\tilde{s}_3,\tilde{s}_4]^{T} =\mathbb{G}_1
[\tilde{s}_1,\tilde{s}_2]^{T} +
\mathbb{G}_2[\tilde{s}_3,\tilde{s}_4]^{T}
\]
Thus, $\mathbb{G}= [\mathbb{G}_1\mid \mathbb{G}_2]\in
\mathbb{R}^{8\times 8}$ is the generator matrix of the code.
Specifically, $\mathbb{G}_1 \in \mathbb{R}^{8\times 4}$ is the
generator matrix of $\mathbf{X}_{1,2}$, and $\mathbb{G}_2 \in
\mathbb{R}^{8\times 4}$ is the generator matrix of
$\mathbf{X}_{3,4}$. The matrix $\mathbb{G}_1$ has the structure
of~(\ref{generator1}) with coefficients $\alpha_{1,2}$ and
$\beta_{1,2}$:
\begin{equation}\label{R1}
 \mathbb{G}_1 \triangleq \left[\mathbf{g}_1\mid \mathbf{g}_2\mid
 \mathbf{g}_3\mid \mathbf{g}_4\right]\triangleq\left[\begin{array}{cc}
 \check{\alpha}_{1,2} & \check{0}\\
 \check{0} & \check{\alpha}_{1,2}\\
 \check{0} & \bar{\check{\beta}}_{1,2}\\
 -\bar{\check{\beta}}_{1,2} & \check{0}
\end{array} \right]
\end{equation}
and
\begin{equation}
 \mathbb{G}_2 \triangleq \left[\mathbf{g}_5\mid
\mathbf{g}_6\mid
 \mathbf{g}_7\mid \mathbf{g}_8\right]\triangleq \left[\begin{array}{rr}
 \check{\varphi}_1  & -\check{\varphi}^{*}_2 \\
 -\check{\varphi}_2 & -\check{\varphi}^{*}_1 \\
 \bar{\check{\varphi}}^{*}_2 & \bar{\check{\varphi}}_1\\
 \bar{\check{\varphi}}^{*}_1 & -\bar{\check{\varphi}}_2
\end{array} \right]
\end{equation}
Direct computation shows that:
\begin{myproperty}
 {\bf\em (Column orthogonality)}~~Both $\mathbb{G}_1$ and
 $\mathbb{G}_2$ have orthogonal columns:
$\langle \mathbf{g}_i, \mathbf{g}_j \rangle = 0$, where $i,j\in
[1,4]$ or $i,j\in [5,8]$, i.e.,
${\mathbb{G}_1}^{\dag}\mathbb{G}_1={\mathbb{G}_2}^{\dag}\mathbb{G}_2=\mathbf{I}_4$.
$\hfill\square$
\end{myproperty}
\begin{myproperty}
{\bf\em (Mutual column orthogonality and Cubic Shaping)}~~With
$\mathbf{T}$ as in~(\ref{TUZ}), the subspace spanned by the columns
of $\mathbb{G}_2$ is orthogonal to the one spanned by the columns of
$\mathbb{G}_1$, i.e., $\langle \mathbf{g}_i, \mathbf{g}_j \rangle =
0$, for $i\in [1,4]$ and $j\in [5,8]$. Since $\mathbb{G} =
[\mathbb{G}_1\mid \mathbb{G}_2]$, we have
\[
\mathbb{G}^{\dag}\mathbb{G} = \left[
                                \begin{array}{cc}
                                  {\mathbb{G}_1}^{\dag}\mathbb{G}_1 & \mathbf{0} \\
                                  \mathbf{0} & {\mathbb{G}_2}^{\dag}\mathbb{G}_2 \\
                                \end{array}
                              \right] =\mathbf{I}_{8}
\]
This implies cubic shaping \cite{FNT}. $\hfill\square$
\end{myproperty}

The matrix $\mathbf{U}$ should be chosen so as to achieve full
rank and maximize the minimum determinant. The best known code of
the form~(\ref{transmitmatrix}) was first found
in~\cite{Tirkkonen}, and independently rediscovered
in~\cite{paredes} and~\cite{FitzIsit07} by numerical optimization.

{\bf Family II}:~~In the second family of fast-decodable
STBCs~\cite{Sari}, both $\mathbf{X}_{1,2}(s_1,s_2)$ and
$\mathbf{X}_{3,4}(s_3,s_4)$ have the Alamouti
structure~(\ref{Alamouti_Structure}), with coefficients
$\alpha_{1,2}, \beta_{1,2}$ used for $\mathbf{X}_{1,2}(s_1,s_2)$,
and $\alpha_{3,4}, \beta_{3,4}$ for $\mathbf{X}_{3,4}(s_3,s_4)$.
The only difference between Family II and Family I is that Family
II codes do not satisfy Property~2: $\mathbb{G}$ is not an
orthogonal matrix, and hence codes in this family exhibit no cubic
shaping.

Table~\ref{mytable} compares the minimum determinant
$\delta_{\min}$ of the best known STBCs in the two families with
that of the Golden code \cite{Golden05} for $4$-, $16$-, and
$64$-QAM signaling. In our computations, we assume that the
constellation points have odd-integer coordinates. It can be seen
that the minimum determinant of Family-I STBCs and of the Golden
code~\cite{Golden05} are constant across constellations, while the
minimum determinant of Family-II STBC decreases slowly as the size
of the signal constellation increases.
The codes of~\cite{Tirkkonen,paredes,FitzIsit07} exhibit a minimum
determinant slightly larger than those of~\cite{Sari}.

\begin{figure}[t]
\begin{center}
\psfig{file=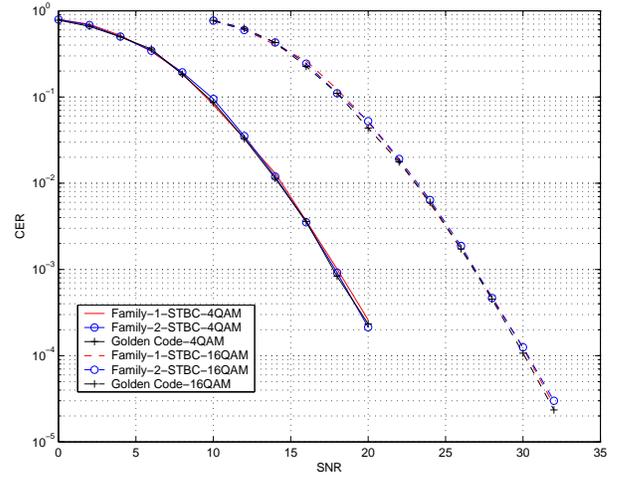,width=80mm,height=64mm}
\end{center}
\caption{Comparison of the CER of the best $2\times 2$ codes in two
fast-decodable STBC families and of the Golden code with $4$- and
$16$-QAM signalings. \label{Fig:compareF1F2GC2x24QAM16QAM}}
\end{figure}

Let us define the signal-to-noise ratio SNR$\triangleq n_t
E_s/N_0$, where $E_s$ the average energy.
Fig.~\ref{Fig:compareF1F2GC2x24QAM16QAM} compares the codeword
error rate (CER) of the best STBCs in the two families and of the
Golden code with $4$- and $16$-QAM signaling. It is shown that
both families of fast-decodable STBCs exhibit similar CER
performances, and both differ slightly, at high SNR, from that of
Golden code. Since the latter has the best CER known, but does not
admit simplified decoding, this small difference can be viewed as
the penalty to be paid for complexity reduction.

\subsection{Decoding Family-I and II STBCs}\label{sec:decoding22}
By direct computation, we have $\langle \mathbf{f}_2,
\mathbf{e}_1\rangle =0$ and $\langle\mathbf{f}_4,
\mathbf{e}_3\rangle=0$. In fact we can see that the full-rate
fast-decodable STBCs are obtained by linearly combining two rate-1
codes: ${\mathbf X}_{1,2}$ and ${\mathbf X}_{3,4}$. Moreover, by
examining the structures of the $2\times 2$ STBCs and the matrix
$\mathbf R$, we obtain the results that follow:

\begin{myproposition}\label{Prop:AlamoutiSTBC} {We have $\langle\mathbf{f}_2,\mathbf{e}_1
\rangle = 0$ if and only if $\mathbf{X}_{1,2}$ is an Alamouti STBC.
Consequently, the fast-decodable full-rate $2\times 2$ STBCs only
exist for $k'=2$ and their corresponding worst-case decoding
complexity does not exceed $2M^3$. $\hfill\square$
}\end{myproposition}

{\bf Proof}: First, if $\mathbf{X}_{1,2}$ is an Alamouti STBC,
from~(\ref{Alamoutistructure}) we conclude that
$\langle\mathbf{f}_2,\mathbf{f}_1 \rangle = 0$, and therefore
\[
\langle\mathbf{f}_2,\mathbf{e}_1 \rangle =
\langle\mathbf{f}_2,\frac{\mathbf{f}_1}{\|\mathbf{f}_1 \|} \rangle =
0
\]
Second, since ${\mathbf X}_{1,2}$ is a rate-1 STBC, it was shown
in~\cite[Theorem 5.4.2]{TarokhSTBC} that complex linear-processing
orthogonal designs only exist in $2$ dimensions and the Alamouti
scheme is unique. Thus, 1) the orthogonality condition $\langle
\mathbf{f}_2,\mathbf{e}_1 \rangle=0$ in $2\times 2$ STBCs implies
that ${\mathbf X}_{1,2}$ must have an Alamouti structure, which
completes the proof of the converse implication; and 2) this also
implies that it is only possible to have $\langle{\mathbf f}_2,
{\mathbf e}_1 \rangle=0$ for the fast-decodable full-rate $2\times
2$ STBCs. Based on Definition~\ref{Def:fastdecSTBC}, it yields
$k'=2$ and the worst-case decoding complexity of $2M^3$.
$\hfill\square$

To further save computational complexity, we may require
$\langle\mathbf{f}_4,\mathbf{e}_3 \rangle = 0$. This can be obtained
if both $\mathbf{X}_{1,2}$ and $\mathbf{X}_{3,4}$ have the Alamouti
structure. Note that this condition is sufficient but not necessary,
since the Alamouti structure implies
$\langle\mathbf{f}_4,\mathbf{e}_3 \rangle = 0$,
but the converse is not true.

The Alamouti structure of $\mathbf{X}_{1,2}$ and $\mathbf{X}_{3,4}$
yields some zero entries in matrix $\mathbf R$ and we have the
following:
\begin{myproposition}\label{Prop:Rmatrix} {The other elements in the
matrix $\bf R$ cannot be nulled. \hfill$\square$}\end{myproposition}

{\bf Proof}: By direct computation we easily verify $\langle
\mathbf{f}_i,\mathbf{f}_j \rangle \neq 0$, $i\in[1,2]$, $j\in
[3,4]$. Therefore this code is not an orthogonal STBC
\cite{TarokhSTBC}, and we have
\begin{equation}
\label{nonorthognal1}\langle\mathbf{f}_3,\mathbf{e}_1 \rangle =
\langle \mathbf{f}_3,\frac{\mathbf{f}_1}{\|\mathbf{f}_1\|} \rangle
\neq 0~~~~\text{and}~~~~\langle\mathbf{f}_4,\mathbf{e}_1 \rangle
\neq 0
\end{equation}
With $\langle {\mathbf f}_2, {\mathbf e}_1\rangle =0$, we have
\begin{equation} \mathbf{e}_2 =
\frac{\mathbf{f}_2 - {\rm Proj}_{\mathbf{e}_1} \mathbf{f}_2}{\Vert
\mathbf{f}_2 - {\rm Proj}_{\mathbf{e}_1} \mathbf{f}_2 \Vert} =
\frac{\mathbf{f}_2}{\Vert \mathbf{f}_2 \Vert}\end{equation} then,
\begin{equation}  \label{nonorthognal2}\langle\mathbf{f}_3,\mathbf{e}_2 \rangle
=\langle\mathbf{f}_3,\frac{\mathbf{f}_2}{\Vert \mathbf{f}_2 \Vert}
\rangle \neq 0~~~~\text{and}~~~~ \langle\mathbf{f}_4,\mathbf{e}_2
\rangle \neq 0\end{equation}
Due to (\ref{nonorthognal1}) and (\ref{nonorthognal2}), the
corresponding elements in $\mathbf R$ cannot be nulled.
$\hfill\square$

In summary, a $2\times 2$ STBC of the form~(\ref{transmitmatrix})
has complexity $2M^3$ if it satisfies
Proposition~\ref{Prop:AlamoutiSTBC}. If in addition ${\mathbf
X}_{3,4}$ has Alamouti structure, then extra computational savings
are available in the SD algorithm. Moreover, if cubic shaping is
required, the generator matrix ${\mathbb G}$ of the STBC is
orthogonal.


\section{New $4\times 2$ STBC and its decoding complexity}\label{sec:newSTBC42}
Here we design a fast-decodable full-rate $(R=2)$ $4\times 2$ STBC
based on the concepts elaborated upon in the previous sections.
Specifically, using the twisted structure described above, we
combine linearly two rate-$1$ codes. Since rate-$1$ orthogonal codes
do not exists for $4$ transmit antennas, we resort quasi-orthogonal
STBCs instead~\cite{JafarSTBC}.
\begin{mydefinition}
{\bf\em (Quasi-orthogonal structure)}~\cite{JafarSTBC}~~A code
whose words have the form
\[
 \mathbf{X} = \left[
\begin{array}{rrrr}
s_1 & -s^{*}_2 & -s^{*}_3 & s_4\\
s_2 & s^{*}_1  & -s^{*}_4 & -s_3\\
s_3 & -s^{*}_4 & s^{*}_1  & -s_2\\
s_4 & s^{*}_3  & s^{*}_2  & s_1
\end{array}%
\right]
\]
or another equivalent form as defined in~\cite{JafarSTBC}, where
$s_i \in \mathbb{C}$, $i=1,\ldots,4$, is said to have a {\em
quasi-orthogonal} structure. The quasi-orthogonal STBC is not full
rank and has $r=2$. $\hfill\square$
\end{mydefinition}
\begin{mydefinition}
{\bf\em (Full-rate, fast-decodable STBC for $4\times 2$ MIMO)} A
full-rate $(R=2)$, fast-decodable STBC for $4\times 2$ MIMO, denoted
${\EuScript G}^\prime$, has $\kappa = 8~{\rm symbols/codeword}$, and
can be decoded by a $12$-dimensional real SD algorithm (rather than
the standard 16-dimensional SD). $\hfill\square$
\end{mydefinition}

The $4\times 4$ codeword matrix $\mathbf{X}\in {\EuScript G}'$
encodes eight QAM symbols $\mathbf{s}=[s_1,\ldots, s_8] \in
\mathbb{Z}^{8}[j]$, and is transmitted by using the channel four
times, so that $T=4$. We admit the sum structure:
\begin{equation}
\mathbf{X} = \mathbf{X}_{1,2}(s_1,s_2,s_3,s_4) +
\mathbf{X}_{3,4}(s_5,s_6,s_7,s_8)\label{new42matrix}
\end{equation}
where $ \mathbf{X}_{1,2}(s_1,s_2,s_3,s_4)$ is a quasi-orthogonal
STBC, and
\begin{equation}
\mathbf{X}_{3,4}(z_1,z_2,z_3,z_4) \triangleq \mathbf{T}\left[
\begin{array}{rrrr}
z_1 & -z^{*}_2 & -z^{*}_3 & z_4\\
z_2 & z^{*}_1  & -z^{*}_4 & -z_3\\
z_3 & -z^{*}_4 & z^{*}_1  & -z_2\\
z_4 & z^{*}_3  & z^{*}_2  & z_1
\end{array}%
\right]\label{new42XBmatrix}
\end{equation}
with
\begin{equation}
\mathbf{T} = \left[
\begin{array}{rrrr}
1 & 0 & 0 & 0\\
0 & 1 & 0 & 0\\
0 & 0 & -1 & 0\\
0 & 0 & 0 & -1
\end{array}%
\right]\label{new42Tmatrix}
\end{equation}
and
\begin{equation}
\left[z_1,z_2,z_3,z_4\right]^T = \mathbf{U}\left[s_5,s_6,s_7,s_8
\right]^T
\end{equation}
where $z_i \in \mathbb{C}$, $i=1,\ldots,4$, $s_k \in \mathbb{Z}
[j]$, $k=5, \ldots, 8$, and $\mathbf{U}$ is a $4\times 4$ unitary
matrix.

\begin{myremark}
 {\bf\em (Rank 2)} Since the matrix $\mathbf{X}_{1,2}$ has the quasi-orthogonal structure,
 the code does not have full rank. In particular, it has $r=2$. $\hfill\square$
\end{myremark}

\begin{myremark}
{\bf\em (Cubic shaping)} Direct computation shows that the matrix
$\mathbf{T}$ guarantees cubic shaping. $\hfill\square$
\end{myremark}

We conduct a search over the matrices $\mathbf{U}$, leading to the
minimum of $\sum_{\delta} A(2,\delta)$, where the terms
$A(2,\delta)$ represent the total number of pairwise error events
of rank $2$ and product distance $\delta$. Since an exhaustive
search through all $4\times 4$ unitary matrices is too complex, we
focus on those with the form
\begin{equation} \label{UequalDF}
\mathbf{U}=\mathbf{D}\mathbf{P}
\end{equation}
where $\mathbf{P}\triangleq [\exp(j 2\pi \ell n/4) ]$ is a
$4\times 4$ discrete Fourier transform matrix, $\mathbf{D}= {\rm
diag}(\exp(j 2\pi n_\ell/N))$ for some integer $N$, and $n_\ell\in
\{ 0, 1, \ldots, N \}$ for $\ell=1,\ldots, 4$.

For $4$-QAM signaling, taking $N=7$ and $n_\ell= 1,2,5,6$, we have
obtained
\[
\begin{tiny}
\mathbf{U}= \left[
\begin{array}{rrrr}
0.31 + 0.39i &  0.31 + 0.39i &  0.31 + 0.39i  & 0.31 +
0.39i\\
  -0.11 + 0.49i & -0.49 - 0.11i &  0.11 - 0.49i  & 0.49 +
  0.11i\\
  -0.11 - 0.49i  & 0.11 + 0.49i & -0.11 - 0.49i &  0.11 +
  0.49i\\
   0.31 - 0.39i & -0.39 - 0.31i & -0.31 + 0.39i &  0.39 + 0.31i
\end{array}
\right]
\end{tiny}
\]
which yields the minimum $\sum_{\delta} A(2,\delta)$.

\begin{figure}[t]
\vspace{-6mm}
\begin{center}
\psfig{file=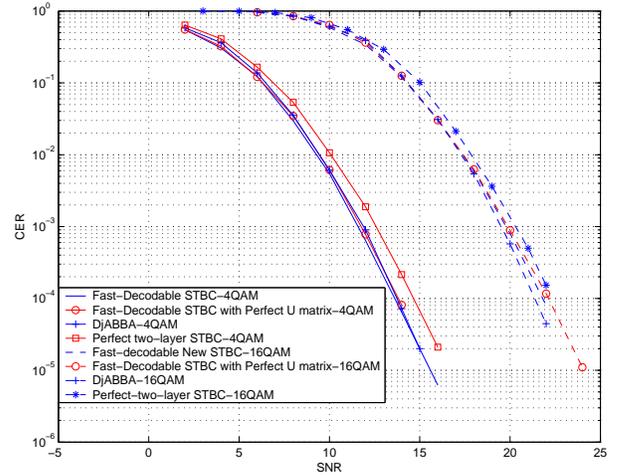,width=80mm,height=64mm}
\end{center}
\caption{Comparison of the CER of different $4\times 2$ STBCs with
4-QAM signaling. \label{Fig:compare42STBC416QAM}}
\end{figure}

Under $4$-QAM signaling, we compare the minimum determinants
$\delta_{min}$ and their associated multiplicities
$A(r,\delta_{min})$, as well as the CERs of the above STBC to the
following $4\times 2$ codes:
\begin{enumerate}
\item Code with the structure~(\ref{new42matrix}),
with $\mathbf{U}$  the $4\times 4$ ``perfect'' rotation
matrix~\cite{Perfect}.
\item The best DjABBA code of~\cite{Hottinen}.
\item The ``perfect''  two-layer code of~\cite{Perfecttwolayer}.
\end{enumerate}
Determinant and multiplicity values are shown in
Table~\ref{table2}. It can be seen that the proposed $4\times 2$
STBC has the smallest $\sum A(2,\delta)$, when compared to the
rank-2 code with perfect rotation matrix $\mathbf U$ in
~\cite{Perfect}. The CERs are shown in
Fig.~\ref{Fig:compare42STBC416QAM}. The proposed code achieves the
best CER up to the CER of $10^{-5}$. Due to the diversity loss,
the performance curves of the new code and the one of DjABBA cross
over at CER of $2\times 10^{-5}$.

For 16-QAM signaling, the best matrix $\mathbf{U}$ with $N=17$ and
$n_\ell= 3,4,5,13$ is
\[
\begin{tiny}
 \mathbf{U}=\left[
\begin{array}{rrrr}
  0.22 + 0.44j &  0.22 + 0.44j &  0.22 + 0.44j  & 0.22 + 0.44j\\
  0.05 + 0.50j & -0.49 + 0.05j &  -0.05 - 0.50  & 0.50 - 0.05j\\
 -0.30 - 0.40j &  0.30 + 0.40j & -0.30 - 0.40j  & 0.30 + 0.40j\\
  0.05 - 0.50j & -0.50 - 0.05j & -0.05 + 0.50j  & 0.50 + 0.05j
\end{array}
\right]
\end{tiny}
\]
The performance of this code is compared with that of other codes in
Fig.~\ref{Fig:compare42STBC416QAM}. We can see that, at CER$=
10^{-4}$, it requires an SNR $0.4$~dB higher than the best known
code of~\cite{Hottinen}, which was not designed for
reduced-complexity decoding.

Finally, we notice that the first two colums of ${\mathbf X}_{1,2}$
are two stacked Alamouti blocks. This provides the orthogonality condition $\langle
{\mathbf f}_2, {\mathbf e}_1\rangle =0$. Therefore the worst-case decoding complexity of fast-decodable $4\times 2$ STBCs
is $2M^7$, as compared to a standard SD complexity $M^8$.

\section{Conclusion}\label{Concl}
We have derived conditions for reduced-complexity ML decoding, and
applied them to a unified analysis of two families of full-rate
full-diversity $2\times 2$ STBCs that were recently proposed.
Moreover, we have compared their minimum determinant, CER
performance, and shaping property, and examined how both families
allow low-complexity ML decoding. We have also introduced design
criteria of fast-decodable STBCs for $2\times 2$ MIMO. These design
criteria were finally extended to the construction of a
fast-decodable $4\times 2$ code. By combining algebraic and
quasi-orthogonal STBC structures, a new code was found that
outperforms any known $4\times 2$ code for $4$-QAM signaling, yet
with a decoding complexity of $2M^7$ in lieu of the worst-case ML
decoding complexity $M^8$.

\bibliographystyle{IEEE}

\begin{table}
\begin{center}
\begin{tabular}{|c|c|c|c|c|c|} \hline
${\EuScript G}$ & $\delta_{\min}$, 4-QAM & $\delta_{\min}$, 16-QAM
& $\delta_{\min}$, 64-QAM\\ \hline
 1st Family & 2.2857 & 2.2857 & 2.2857\\ \hline
 2nd Family & 1.9973 & 1.9796 & 1.8784\\ \hline
Golden Code & 3.2 & 3.2 & 3.2\\ \hline
 \end{tabular}
\end{center}
 \caption{The minimum determinants $\delta_{\min}$ of the Golden code and two families of
fast-decodable STBCs with $4$-, $16$-, and $64$-QAM signaling.
\label{mytable}}
\end{table}
\begin{table}[t]
\begin{center}
\begin{tabular}{|c|c|c|} \hline
Codes & $\delta_{\rm min}$ & Multiplicities \\
\hline\hline New STBC & $0$ & $\sum_{\delta} A(2,\delta) = 160$\\
\hline\hline Perfect Code $\mathbf{U}$ matrix & $0$ & $\sum_{\delta} A(2,\delta) = 560$ \\
\hline\hline  DjABBA & $0.8304$ & $A(4,0.8304) = 770$\\
\hline\hline  Two-Layers Perfect Code & $0.0016$ & $A(4,0.0016) = 128$\\
\hline
\end{tabular}
\end{center}
\caption{Minimum determinants of $4\times 2$ STBCs with $4$-QAM
signaling \label{table2}}
\end{table}
\newpage

\end{document}